\documentclass{PoS}

\title{The Effects of the Galactic Magnetic Field on UHECR From Local Sources}

\ShortTitle{GMF Effect on UHECR}

\author{\speaker{Andrew~M.~Taylor}\\
  Deutsches Elektronen-Synchrotron (DESY), Platanenallee 6, D-15738 Zeuthen, Germany\\
  E-mail: \email{andrew.taylor@desy.de}}

\author{{A.~M.~Hillas\thanks{Deceased.}}\\
  School of Physics and Astronomy, University of Leeds, Leeds LS2 9JT, UK\\
}

\abstract{We summarise the results found following a study of the effects that 
Galactic magnetic fields can have on the propagation of  cosmic rays 
from local extragalactic sources. This study focuses on the coherent structures
in the Jansson-Farrar Galactic magnetic field model, namely the:  
disk field, a toroidal field, and an x-field components. The phenomena of 
Galactic magnetic field shadowing, source deflection, and Galactic magnetic 
field tunnel vision are all noted. Attention throughout this study is placed on 
particles with rigidity around $10^{18.5}$~EV, believed to dominate the
cosmic ray spectrum above the ankle. The Galactic magnetic field 
component predominantly responsible for giving rise to each of these 
different effects was determined.}

\FullConference{36th International Cosmic Ray Conference -ICRC2019-\\
		July 24th - August 1st, 2019\\
		Madison, WI, U.S.A.}

\begin{document}

\section{Introduction}

% UHECR source requirements
Although the origin of UHECR still remains unknown, a wealth of information continues
to mount about their candidate sources. 

The maximum energy for a nuclei achievable by acceleration within a candidate source~\cite{Hillas:1984},
can be estimated through a balance of the acceleration timescale with either the source activity or 
escape timescales. By expressing the acceleration timescale $t_{\rm acc}$ 
as a multiple $\eta$ of the Larmor time $R_{\rm L}/c=p/(eB)$ of a particle of momentum $p$ 
around the magnetic field (strength $B$): $t_{\rm acc}=\eta R_{\rm L}/c$. For a source's maximum 
energy to reach up to UHECR energies, it must operate as an extreme accelerator, achieving 
close to Bohm level scattering times, with $\eta\approx 1$.

A tight lower bound can be placed on the magnetic luminosity of an UHECR sources,
\cite{1995ApJ...454...60N}, which can be written as 
$L_B\gtrsim 10^{43}\,\eta^2 \left(R_{\rm max}/10^{19}{\rm\, V}\right)^2\,$erg/s, where 
$R = E/Z$ is the rigidity of the UHECRs in the observer frame. A maximum rigidity of 
$10^{19}{\rm\, V}$, that is $E = 10^{19}{\rm\, eV}$ for protons ($Z=1$) or 
$E \sim 10^{20}{\rm\, eV}$ for Silicon ($Z=14$), is in line with the current best 
constraints from the Pierre Auger Observatory \cite{Aab:2014aea, AlvesBatista:2019tlv}.

% Need for local sources
The propagation of UHECR is limited by their interaction with extragalactic 
background radiation fields. A maximum distance scale of 80~Mpc has been previously 
determined, within which the nearest by UHECR sources must exist \cite{Taylor:2011ta}. 
Although no blazars sit within this range, several examples of their misaligned 
counterparts (radiogalaxies) do satisfy this local proximity constraint.

% Local misaligned blazars
As an example of such objects, local radiogalaxy such as Cen~A exist, whose jet power
appears to sufficient in order reach to UHECR energies. The jet power of Cen~A being of 
the order of $10^{43}\,$erg/s \cite{Wykes:2013gba}, the above estimates suggest that this 
local misaligned AGN, and others like it, are capable of accelerating nuclei up to the 
UHECR energy scale, provided $\eta$ is 
close to unity at the highest energies. Hence, whether  blazars and radio galaxies can 
actually contribute to the UHECR spectrum or not, depends both on how efficient the 
acceleration can be, and on the composition of UHECRs themselves. 

% Need to consider propagation from local sources
Of the deflection of UHECR during their propagation in the extragalactic environment, 
the deflection of UHECR from local extragalactic sources will be smallest due to their
reduced residence time within extragalactic magnetic fields. The deflection of UHECR within the Galactic
magnetic field can be of particular relevance for particles originating from local 
sources. Likewise, a sufficiently accurate understanding of the Galactic magnetic field 
structure allows the promise a recovery of the UHECR flux impinging on the Galactic
magnetosphere, whose anisotropy may be dominated by the local source fluxes. 

We here focus on obtaining an understanding of the effects that the Galactic magnetic
field can have on the arriving UHECR. By further considering the components
of the Galactic magnetic field presently considered in a popular Galactic magnetic 
field model, the field components predominantly giving rise to these effects are isolated.

\section{The Galactic Magnetic Field Model}

% Introduction of different coherent structure components in Jansson and Farrar model
The Galactic magnetic field model considered here is the regular component of the
JF12 model \cite{Jansson:2012pc}. This consists of three separate structured field
components, namely: a disk field component ({\bf disk}), a toroidal halo component
({\bf toroidal}), and an x-field component ({\bf x-field}). The field resulting 
from the sum of these components we refer to throughout as {\bf total}.

% Where knowledge about Galactic Magnetic Field structure is derived from
One of the constraints on this proposed Galactic magnetic field structure have been provided 
by Faraday rotation measure measurements from an ensemble of region in the sky. Such
observations constrain the parallel component of the magnetic field to the line-of-sight, 
requiring also the adoption of a Galactic thermal electron distribution in order to
provide such a constraint.

Polarised synchrotron emission maps can provides a complimentary constraint on the Galactic
magnetic field, providing input on the perpendicular component of the field to the 
line-of-sight. The use of synchrotron maps to provide such a constraint, however, 
first require the adoption of a Galactic non-thermal electron distribution.

The inference of the Galactic magnetic field structure is therefore limited by
both our incomplete knowledge of the thermal and non-thermal contents of the Galaxy,
as well as current observational limitations in our ability to probe Galactic rotation 
measure and synchrotron emission.

\section{The Influence of the Galactic Magnetic Field on UHECR}

% Energetics of model
We first consider the energy content of the three Galactic
magnetic field components of the JF12 model. These are $U_{\rm B}^{\rm disk}=8\times 10^{53}$~erg,
$U_{\rm B}^{\rm toroid}=4\times 10^{54}$~erg, and $U_{\rm B}^{\rm x-field}=3\times 10^{54}$~erg.
In comparison, the Galactic CR population is estimated to maintain a total energy 
content of $U_{\rm CR}=3\times 10^{55}$~erg \cite{Drury:2012md}. Furthermore, a consideration
of the spatial distribution of these magnetic field components, suggests that only the
{\bf toroidal} and {\bf x-field} dominate the out-of-plane Galactic magnetic field 
structure, and therefore are potentially of most relevance for the majority of extragalactic 
source.

% Analogy with Heliosphere and its distortion of our perspective on LIS
With the Galactic magnetic field structured components considered possessing sizes larger
than the Larmor radii of $3\times 10^{18}$~eV protons in $\mu G$ strength magnetic
fields, the presence of these structures is naturally expected
to distort the flux passing through the Galaxy. Specifically, both the 
spatial distribution of the cosmic rays throughout the Galaxy, and their 
angular distribution locally at a point within in, will be collectively effected.

\subsection{Shadowing}

We here consider a setup in which UHECR from Cen~A (with Galactic coordinates 
$l=309.5^{\circ}$, $b=19.4^{\circ}$) arrive to the Galactic magnetosphere
in a parallel beam of radius 30~kpc. With an injection distance from the Galactic
center for this beam of 30~kpc being adopted, half of the Galactic magnetosphere 
was illuminated by the beam. 

Following the injection of these $10^{18.5}$~eV protons into the system, their subsequent 
trajectories were tracked until their eventual escape. A Galactic escape boundary of 30~kpc 
was here adopted. Following the assumption that the source
is continuously illuminating the Galactic magnetosphere, the steady-state density in
the Galactic disk region (within 100~pc from the plane) was determined.

\begin{figure}[t!]
\begin{center}
\includegraphics[angle=-90,width=0.49\textwidth]{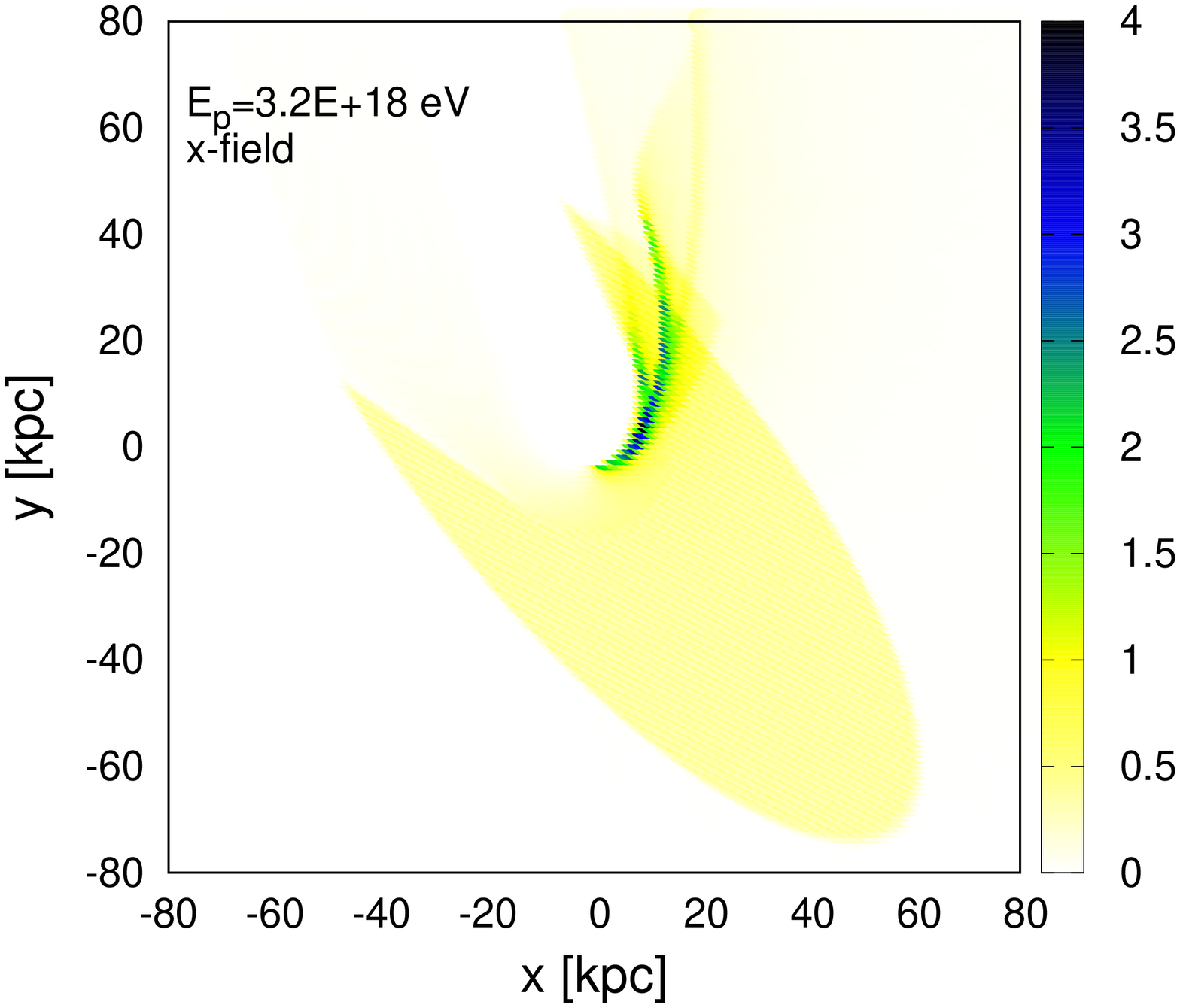}
\includegraphics[angle=-90,width=0.49\textwidth]{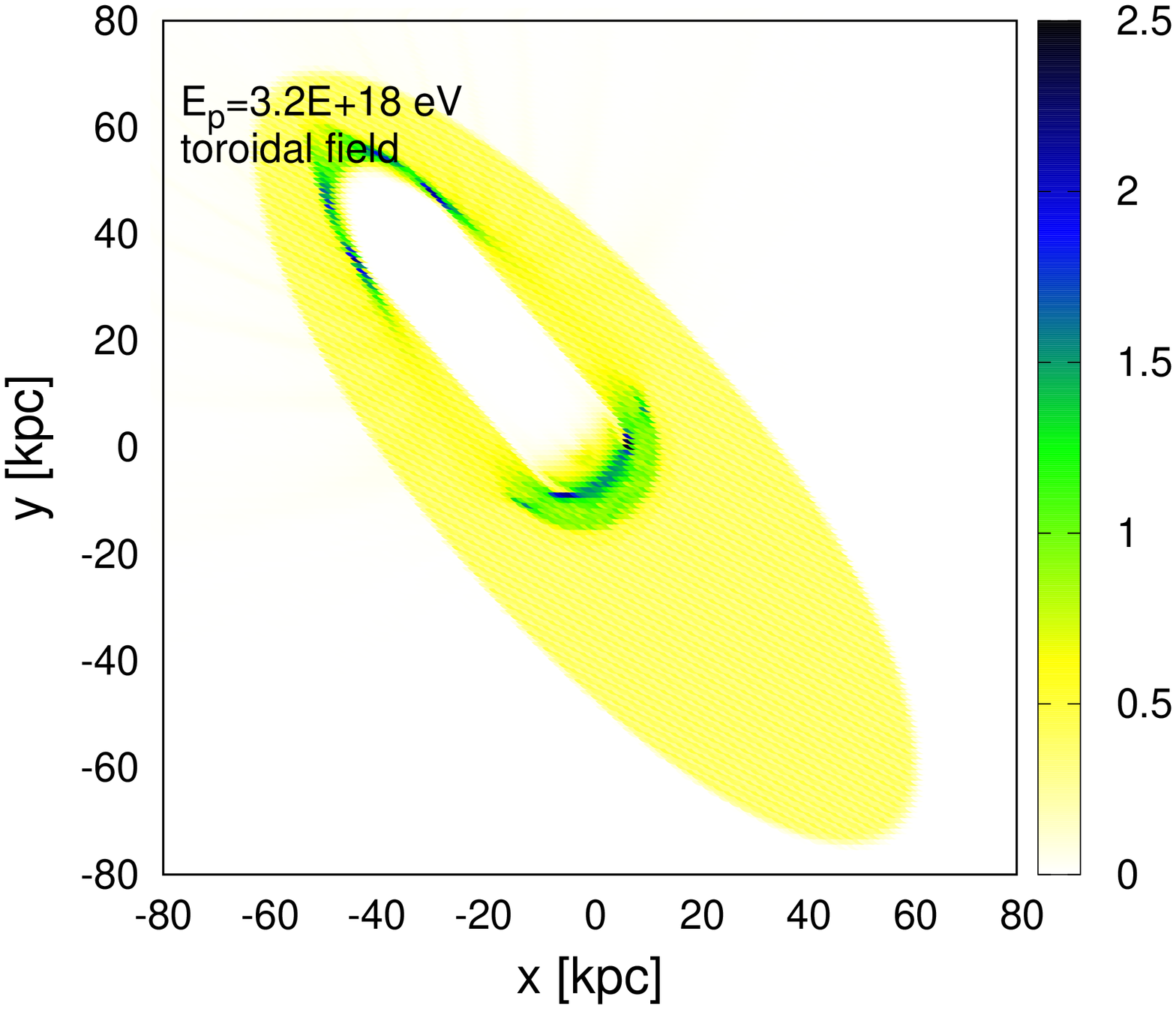}\\
%\begin{center}
\includegraphics[angle=-90,width=0.49\textwidth]{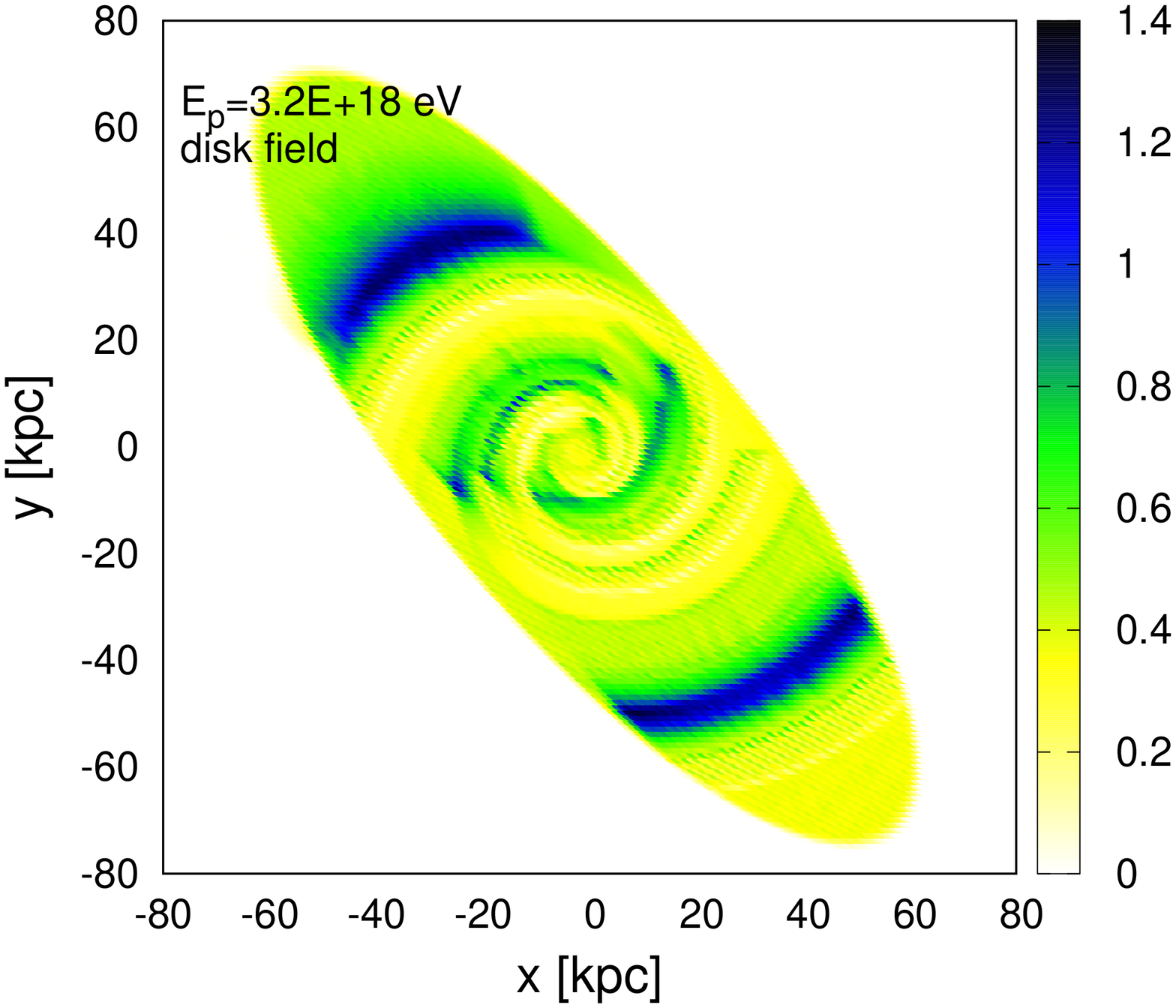}
\includegraphics[angle=-90,width=0.49\textwidth]{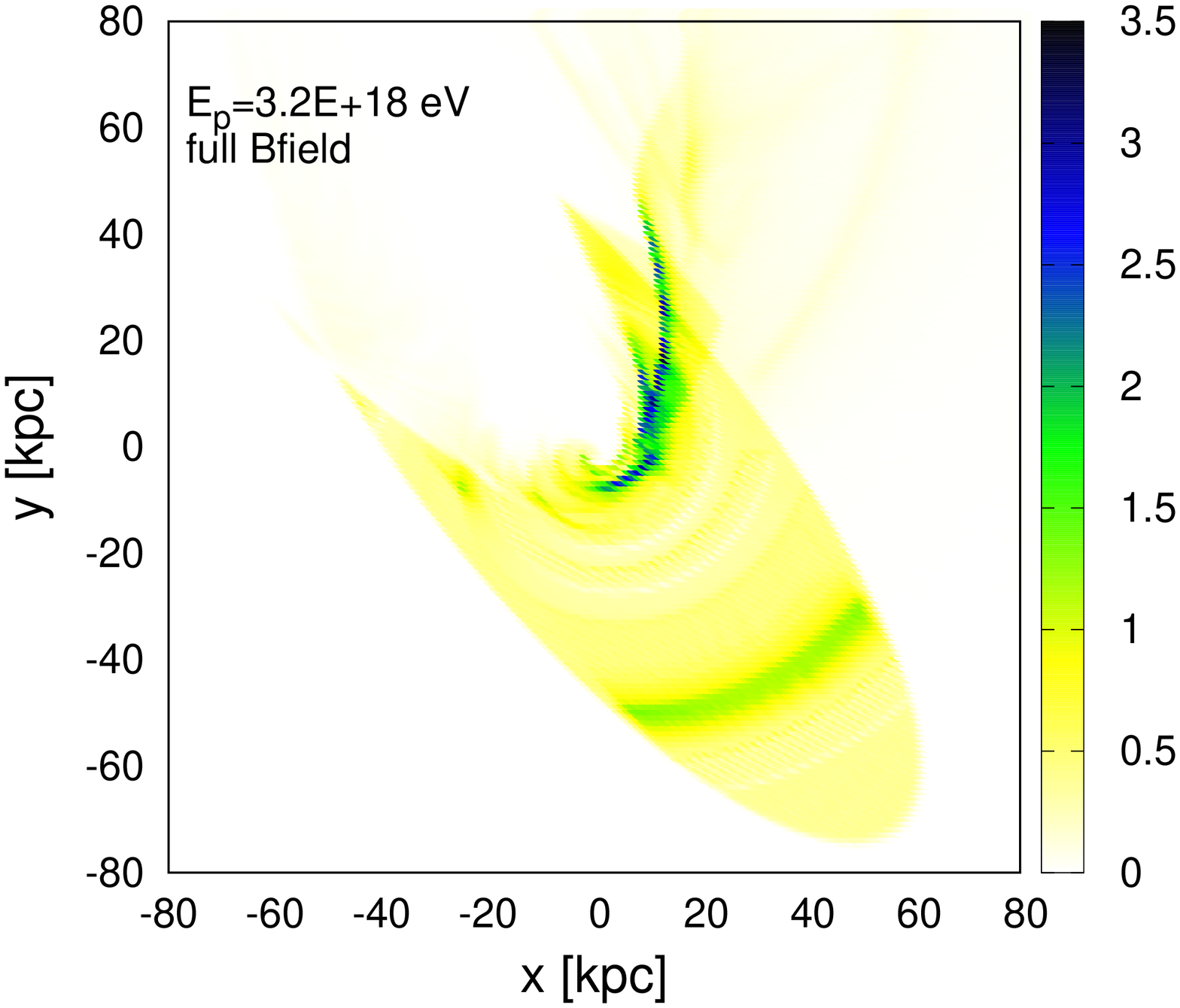}
%\end{center}
\caption{The steady-state CR density in the Galactic disk region following the
continuous injection of $10^{18.5}$~eV protons (see injection setup description in the 
main text). The top-left/top-right/bottom-left/bottom-right plots shows the result for the 
{\bf x-field}/{\bf toroidal}/{\bf disk}/{\bf total} Galactic magnetic field components.}
\label{Shadowing}
\end{center}
\end{figure}

The results for the case of only single component of the Galactic magnetic field,
namely {\bf x-field}, {\bf toroidal}, and {\bf disk} are shown in fig.~\ref{Shadowing}.
Both components whose contribution outside the Galactic disk plane is significant
(the {\bf x-field} and {\bf toroidal} components) are found to lead to a considerable level 
of shadowing of the region on the far side of the Galactic center relative to the
source position. It should be noted that the overall elliptical illumination shape within
the Galactic plane results from the projection effect of the inclined source beam onto
the Galactic plane surface.

The presence of such a Galactic shadowing process is somewhat analogous to \cite{Amenomori:2013own}, 
where the dense magnetic field surrounding the Sun leads to shielding of the Galactic cosmic rays
from a region around the Solar disk. However, in the case considered here, the shadowing
process is instead be being applied to extragalactic CR from a local source, 
following their passing close the Galactic center region.

\subsection{Shifting in Source Position}

\begin{figure}[t!]
\begin{center}
\includegraphics[angle=-90,width=0.49\textwidth]{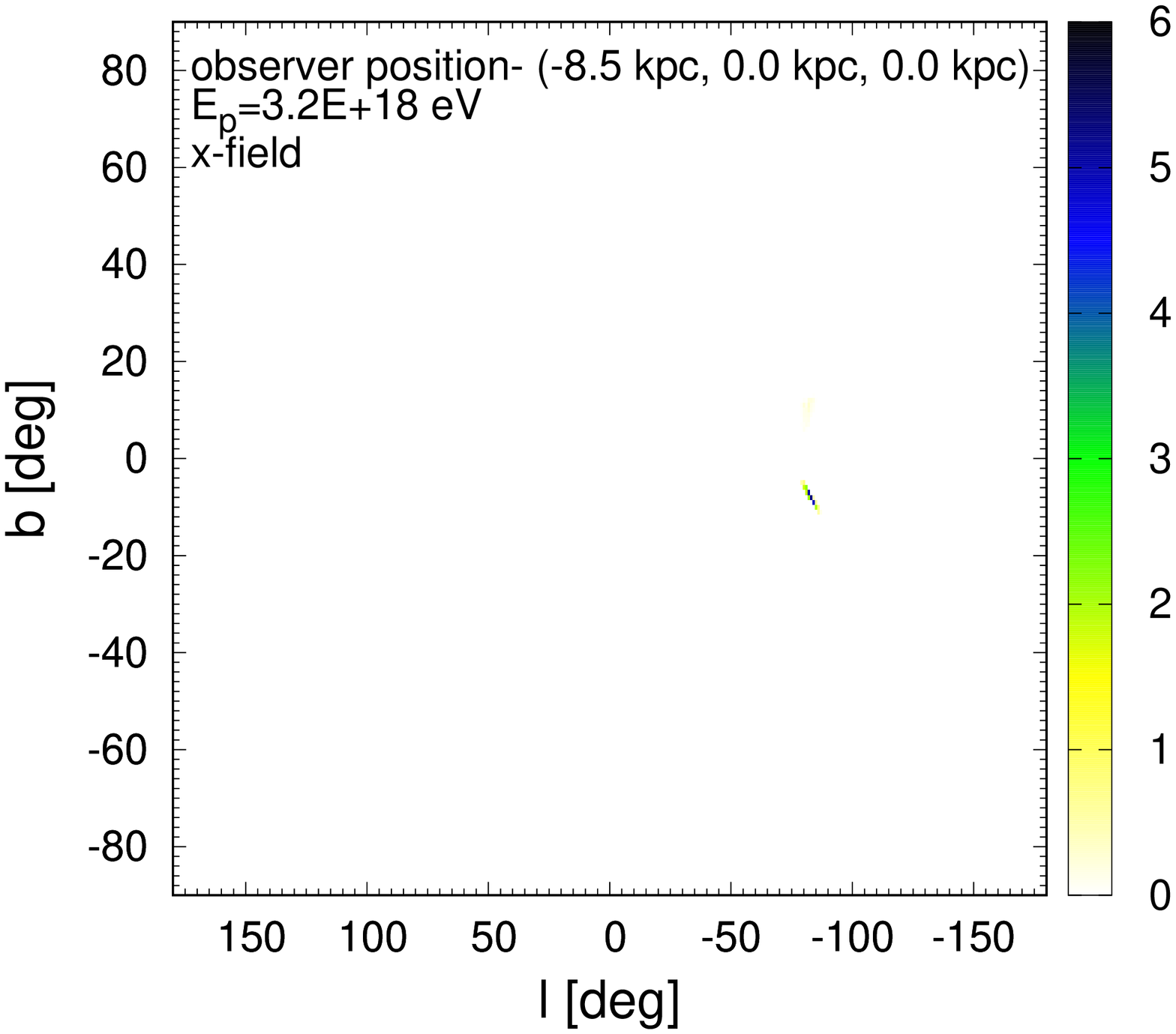}
\includegraphics[angle=-90,width=0.49\textwidth]{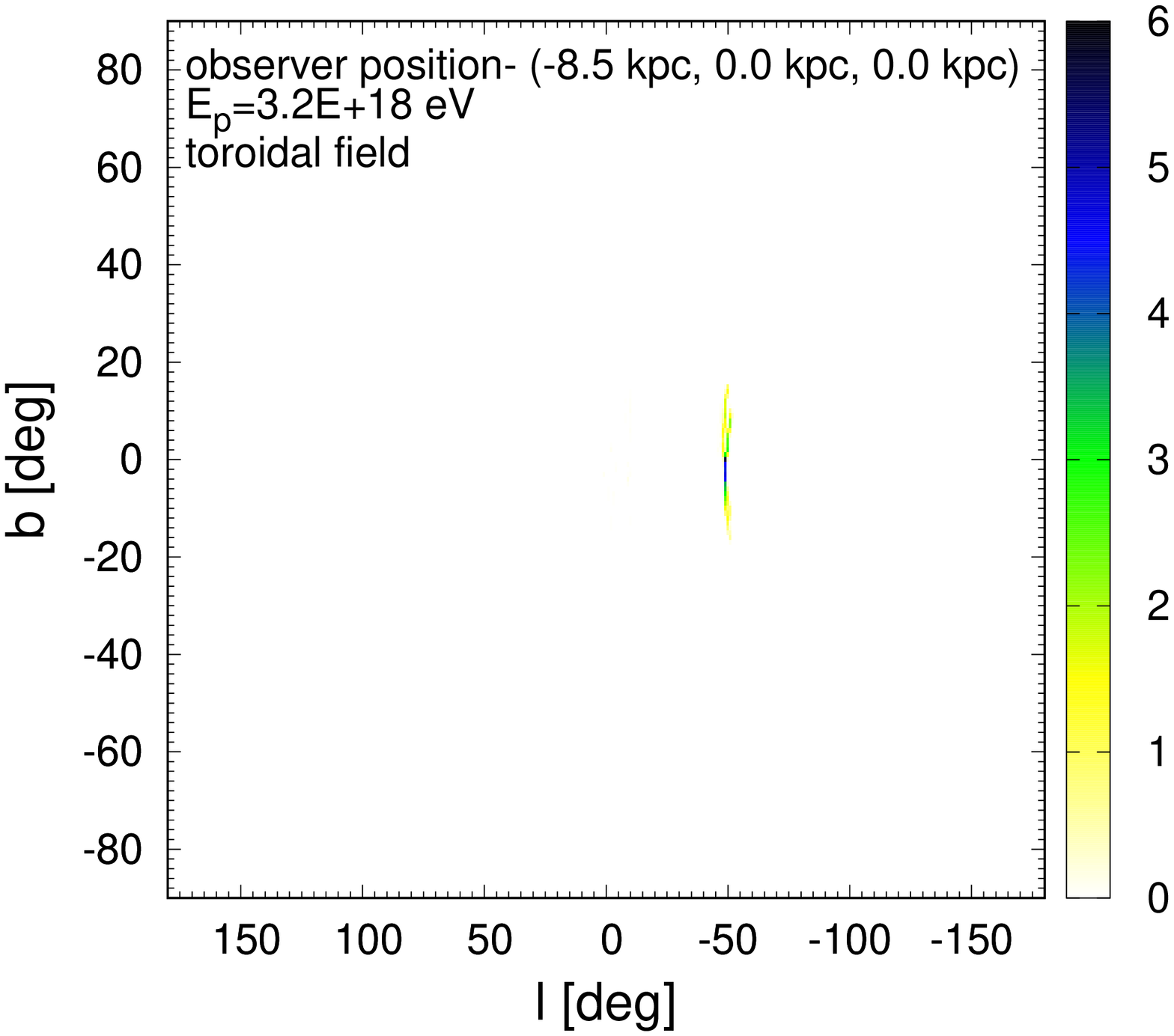}\\
%\begin{center}
\includegraphics[angle=-90,width=0.49\textwidth]{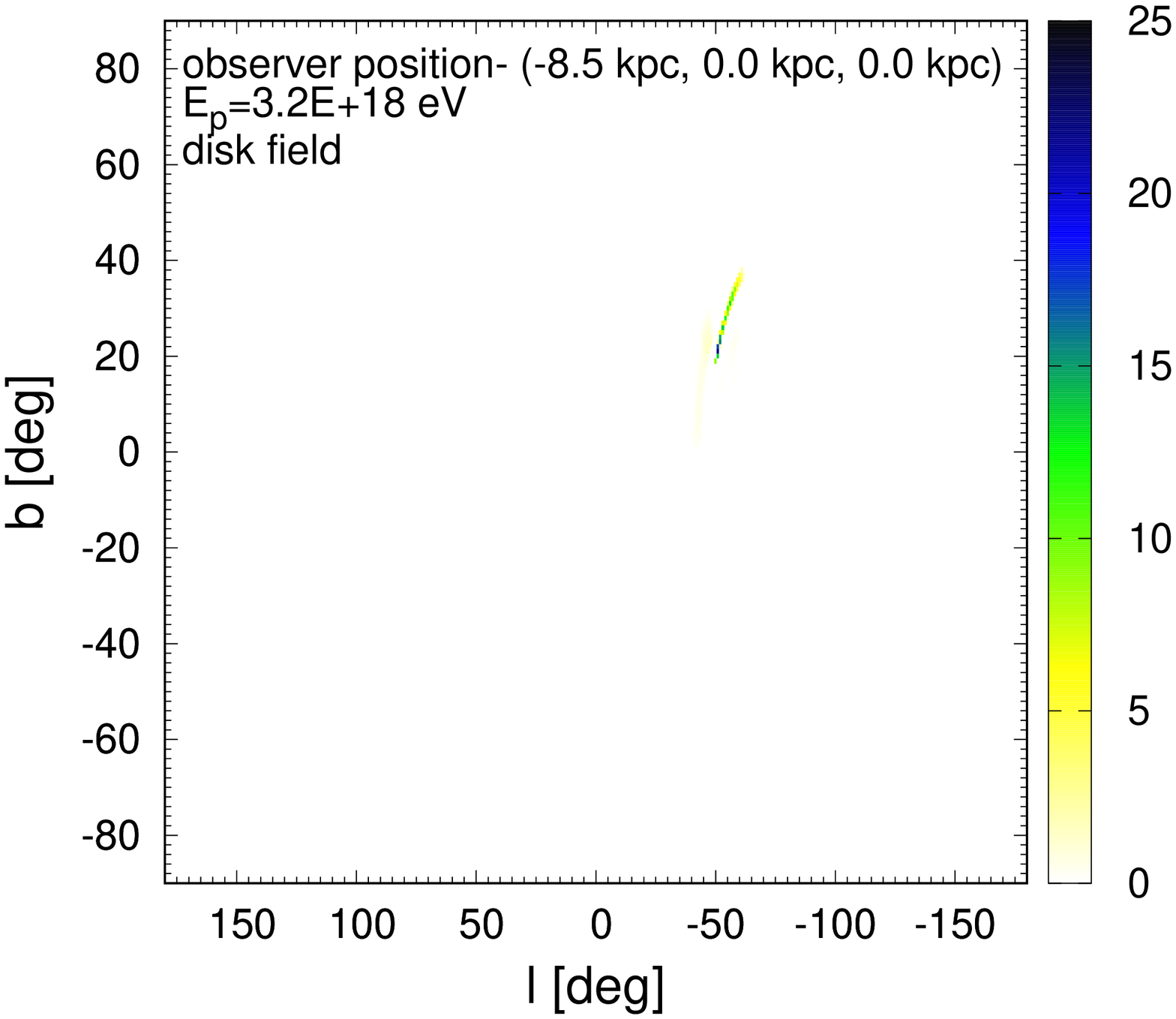}
\includegraphics[angle=-90,width=0.49\textwidth]{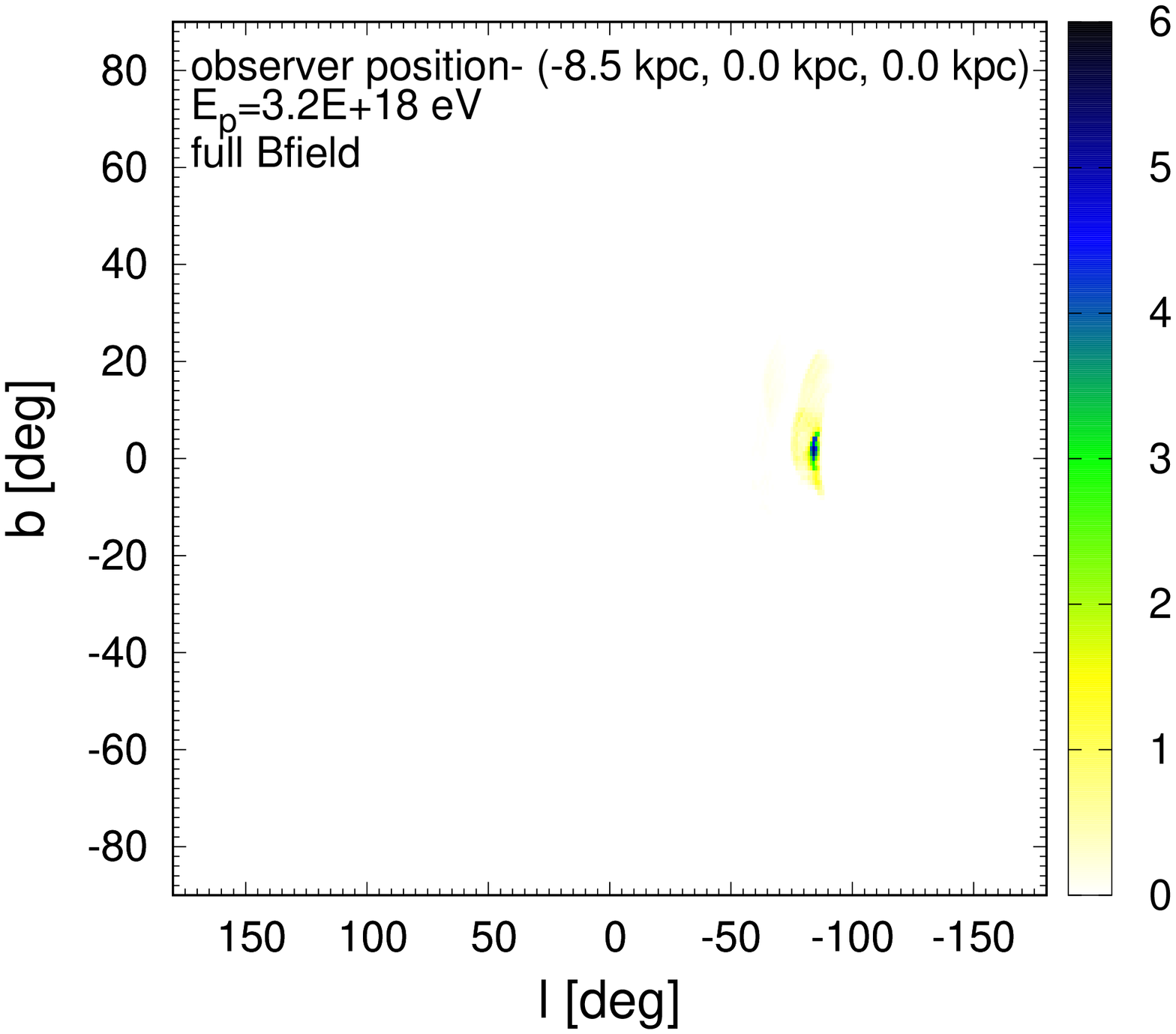}
%\end{center}
\caption{The arriving CR anisotropy skymaps following the injection of $10^{18.5}$~eV
protons  (see injection setup description in the main text). The top-left/top-right/bottom-left/bottom-right plots shows the result for the {\bf x-field}/{\bf toroidal}/{\bf disk}/{\bf total} Galactic magnetic field components.}
\label{Shift}
\end{center}
\end{figure}

Utilising the same setup for  the propagation of CR as described in the previous 
section, the anisotropy of the particles arriving to the Earth location was next 
considered. Although the particle beam recalled its general origin, a degree of 
shifting of the beam position relative to its initial direction was found (see bottom-right
panel in fig. \ref{Shift}).

This overall shift in the source position is consistent with the
finding in \cite{Keivani:2014kua}. It should be noted that although we focus here
on the injection of particles from the source direction, the method of backtracking
particles from Earth's position was also found to give consistent results.

An investigation was next carried out in order to determine which component
of the Galactic magnetic field played the dominant role in the shifting of
the source position. This was carried out by a switching off of the 
{\bf disk}, {\bf torus}, and {\bf x-field} components. The findings are shown
in the top-left, top-right, and bottom-left panels in fig.~\ref{Shift}. 
It is evident from these results that the {\bf x-field} component dominated the coherent 
shifting of the source position, particularly in the Galactic longitude. The removal of 
this field component being found to allow a much broader angular distribution range of 
CR to arrive from Cen~A.

\subsection{Tunnel Vision}

\begin{figure}[t!]
\begin{center}
\includegraphics[angle=-90,width=0.49\textwidth]{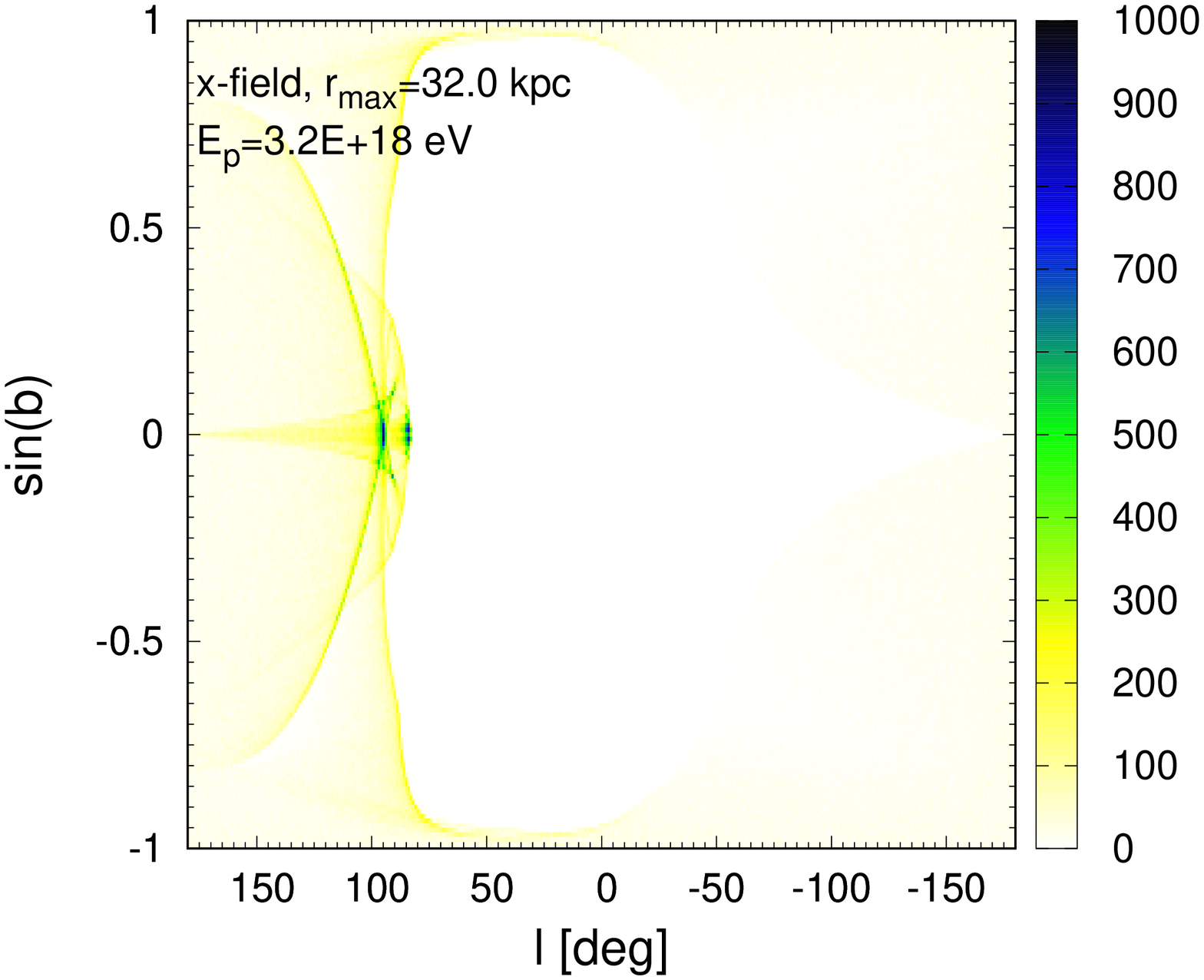}
\includegraphics[angle=-90,width=0.49\textwidth]{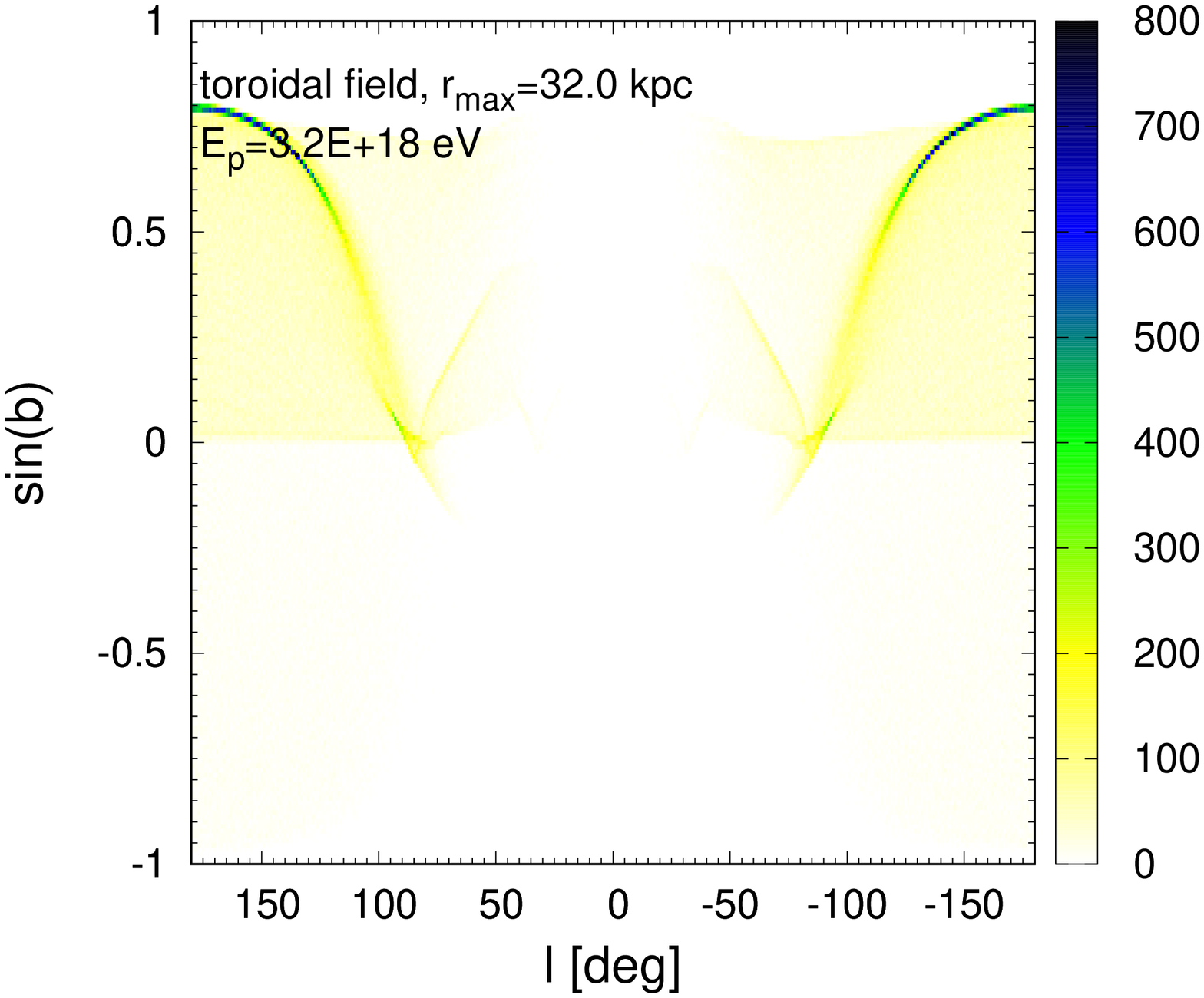}\\
%\begin{center}
\includegraphics[angle=-90,width=0.49\textwidth]{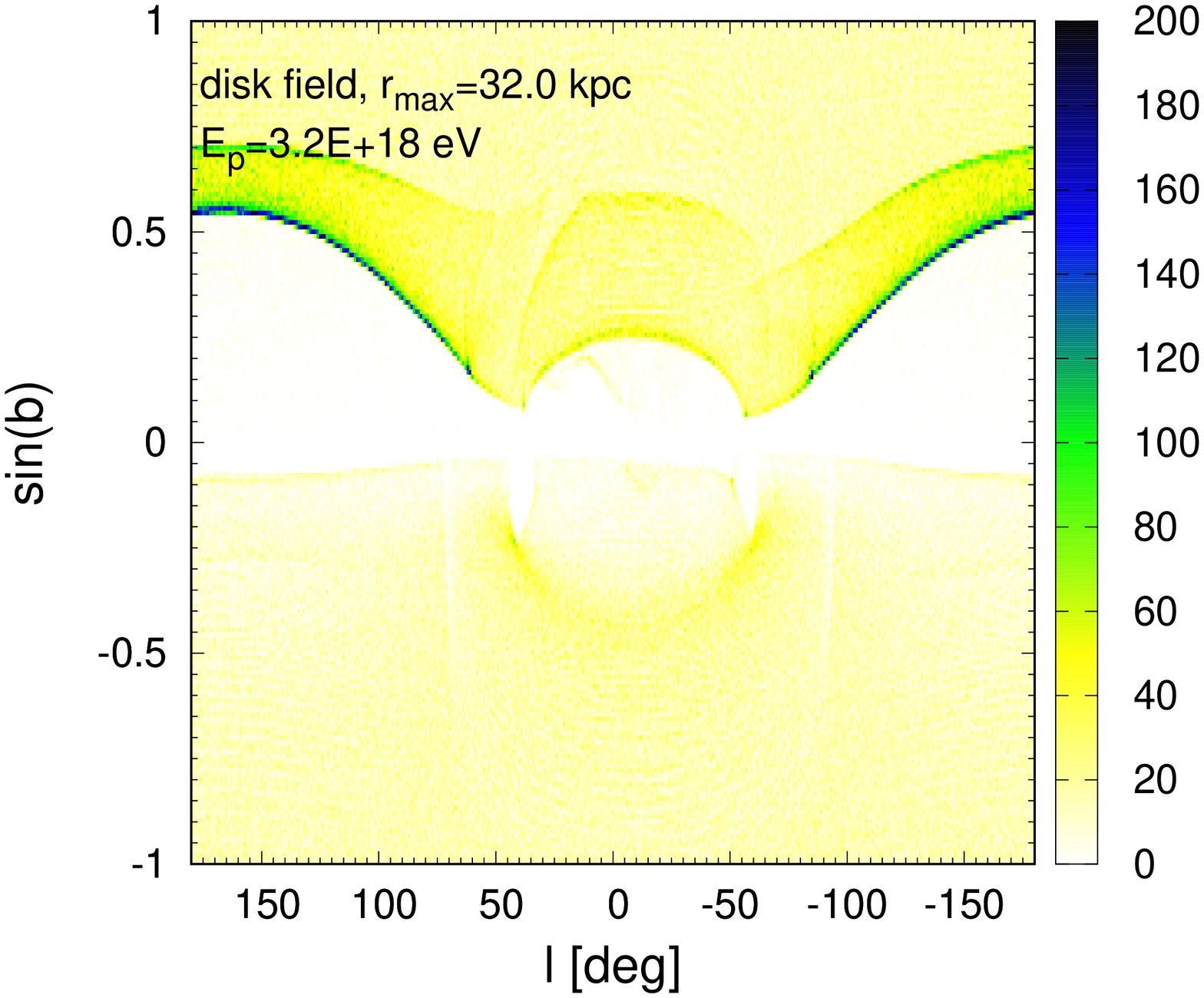}
\includegraphics[angle=-90,width=0.49\textwidth]{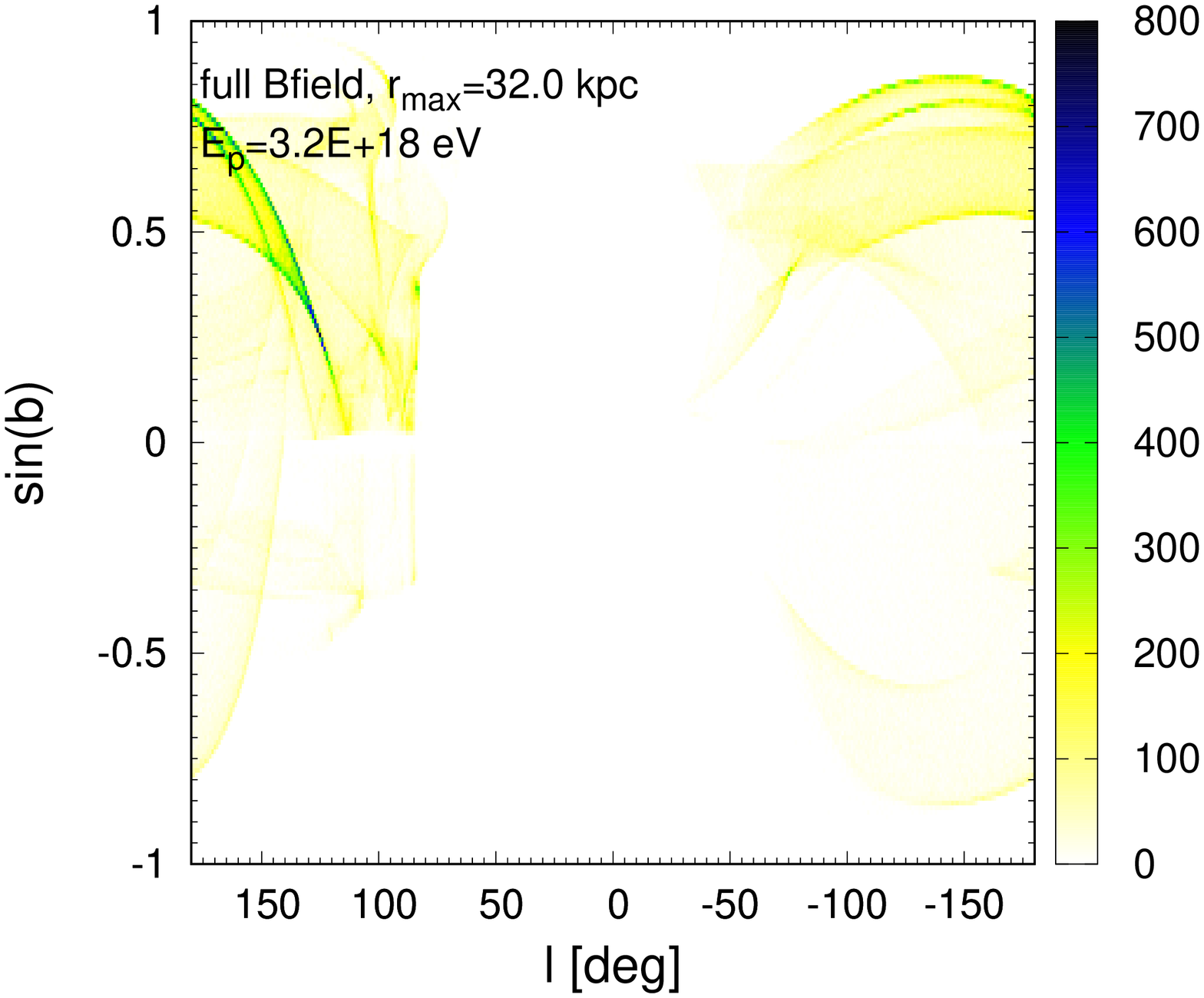}
%\end{center}
\caption{Skymaps of the backtracked CRs following the injection of isotropic distribution
of $10^{18.5}$~eV protons at Earth. The top-left/top-right/bottom-left/bottom-right plots shows the result for the {\bf x-field}/{\bf toroidal}/{\bf disk}/{\bf total} Galactic magnetic field components.}
\label{Tunnel}
\end{center}
\end{figure}

Finally, the part of the extragalactic sky that our present location within the Galaxy 
forces us to preferentially observe was probed. Starting with an isotropic ensemble of 
$10^{18.5}$~eV protons from Earth and backtracking them through the Galactic magnetic field, 
their subsequent anisotropic probing of the extragalactic sky was investigated. The results 
for the backtracking through the different individual components of the Galactic magnetic 
field are shown in fig.~\ref{Tunnel}.

It is apparent from these results that the Northern extragalactic sky (ie. positive Galactic 
latitudes) is preferentially probed in direction at angles point from the Galactic center 
direction. The origin for this preference can be understood from the {\bf toroid} plot in 
fig.~\ref{Tunnel}. This figure demonstrates the {\bf toroid} Galactic magnetic field component
acts as a magnetic lenses, deterring flux from southern Galactic latitudes. The origin of 
the preferential sampling directions at large angles to that of the Galactic center region
is instead found to originate from the presence of both {\bf toroid} and {\bf x-field} 
components.

\section{Conclusion}

Motivated by the need for local extragalactic sources, consideration is placed on how 
the observation of these local sources is altered by the presence of the Galactic magnetic field. 
The effect of the potential coherent Galactic magnetic field structures on the entry of
extragalactic cosmic rays into the Galactic magnetosphere are considered. 

% Shadowing
Shadowing of $10^{18.5}$~eV UHECR passing close to the Galactic center region for trajectories 
originating from Cen~A was found. This result was noted to be influenced by both the 
{\bf toroidal} and {\bf x-field} Galactic magnetic field components, with the latter of the 
two components being found to dominate this shadowing effect.

% Shifting
For $10^{18.5}$~eV UHECR originating from the direction of Cen~A, a shifting of the source 
direction was also seen. This shifting was again found to be influenced by the {\bf toroidal} 
and {\bf x-field} Galactic magnetic field components, with the former of the two being found 
to dominate this source shifting effect in Galactic latitude, and the latter of the two 
dominating the source shifting effect in Galactic longitude.

% Tunnel vision
Lastly, the tunnel vision due to our observation of Galactic UHECR from Earth's location within
the Galactic magnetic field was considered. The parts of the extragalactic sky preferentially 
probed by UHECR observation at Earth was investigated by backtracking of an isotropic distribution 
of UHECR at Earth through the Galactic magnetic field. A preference for observations of the 
northern Galactic hemisphere, and directions away from the Galactic center direction were
found. Again, these effect were found to originate from both the {\bf toroidal} and {\bf x-field} 
Galactic magnetic field components of the JF12 model, with the former of the two components leading 
to the preference of northern hemisphere observation.

With a present very incomplete description of the Galactic magnetic field, the assignment of the
effects of the potential components of the field on the arriving cosmic rays is considered
of benefit. From this study, emphasis is placed on the out of plane Galactic magnetic field 
components, namely the {\bf toroidal} and {\bf x-field} Galactic magnetic field components.
Our limited understanding of these components strongly constrains our ability to accurately
describe the shadowing, source shifting, and tunnel vision effects noted in this study.


\begin{thebibliography}{99}

\bibitem{Hillas:1984} A. M. Hillas,
 %``The origin of ultra-high-energy cosmic rays,''
 Ann. Rev. Astron. Astrophys. {\bf 22}, 425 (1984).


\bibitem{Taylor:2011ta}
  A.~M.~Taylor, M.~Ahlers and F.~A.~Aharonian,
  %``The need for a local source of UHE CR nuclei,''
  Phys.\ Rev.\ D {\bf 84} (2011) 105007
  doi:10.1103/PhysRevD.84.105007
  [arXiv:1107.2055 [astro-ph.HE]].
  %%CITATION = doi:10.1103/PhysRevD.84.105007;%%

\bibitem{1995ApJ...454...60N}
  C.~A.~Norman, D.~B.~Melrose, A.~Achterberg,
  Astrophysical Journal 454 60N (1995)

\bibitem{Aab:2014aea}
  A.~Aab {\it et al.} [Pierre Auger Collaboration],
  %``Depth of maximum of air-shower profiles at the Pierre Auger Observatory. II. Composition implications,''
  Phys.\ Rev.\ D {\bf 90} (2014) no.12,  122006
  doi:10.1103/PhysRevD.90.122006
  [arXiv:1409.5083 [astro-ph.HE]].
  %%CITATION = doi:10.1103/PhysRevD.90.122006;%%

\bibitem{AlvesBatista:2019tlv}
  R.~Alves Batista {\it et al.},
  %``Open Questions in Cosmic-Ray Research at Ultrahigh Energies,''
  Front.\ Astron.\ Space Sci.\  {\bf 6} (2019) 23
  doi:10.3389/fspas.2019.00023
  [arXiv:1903.06714 [astro-ph.HE]].
  %%CITATION = doi:10.3389/fspas.2019.00023;%%

\bibitem{Wykes:2013gba}
  S.~Wykes {\it et al.},
  %``Mass entrainment and turbulence-driven acceleration of ultra-high energy cosmic rays in Centaurus A,''
  Astron.\ Astrophys.\  {\bf 558} (2013) A19
  doi:10.1051/0004-6361/201321622
  [arXiv:1305.2761 [astro-ph.HE]].
  %%CITATION = doi:10.1051/0004-6361/201321622;%%

\bibitem{Jansson:2012pc}
  R.~Jansson and G.~R.~Farrar,
  %``A New Model of the Galactic Magnetic Field,''
  Astrophys.\ J.\  {\bf 757} (2012) 14
  doi:10.1088/0004-637X/757/1/14
  [arXiv:1204.3662 [astro-ph.GA]].
  %%CITATION = doi:10.1088/0004-637X/757/1/14;%%

\bibitem{Drury:2012md}
  L.~O.~Drury,
  %``Origin of Cosmic Rays,''
  Astropart.\ Phys.\  {\bf 39-40} (2012) 52
  doi:10.1016/j.astropartphys.2012.02.006
  [arXiv:1203.3681 [astro-ph.HE]].
  %%CITATION = doi:10.1016/j.astropartphys.2012.02.006;%%

\bibitem{Amenomori:2013own}
  M.~Amenomori {\it et al.} [Tibet ASgamma Collaboration],
  %``Probe of the Solar Magnetic Field Using the "Cosmic-Ray Shadow" of the Sun,''
  Phys.\ Rev.\ Lett.\  {\bf 111} (2013) no.1,  011101
  doi:10.1103/PhysRevLett.111.011101
  [arXiv:1306.3009 [astro-ph.SR]].
  %%CITATION = doi:10.1103/PhysRevLett.111.011101;%%

\bibitem{Keivani:2014kua}
  A.~Keivani, G.~R.~Farrar and M.~Sutherland,
  %``Magnetic Deflections of Ultra-High Energy Cosmic Rays from Centaurus A,''
  Astropart.\ Phys.\  {\bf 61} (2014) 47
  doi:10.1016/j.astropartphys.2014.07.001
  [arXiv:1406.5249 [astro-ph.HE]].
  %%CITATION = doi:10.1016/j.astropartphys.2014.07.001;%%

\end{thebibliography}
\end{document}